\documentclass{iopart}

\usepackage{graphicx}

\jl{6}        
\eqnobysec    

\def\beq{\begin{equation}}
\def\eeq{\end{equation}}
\def\dualp#1{{}^{\ast_{(\hbox{$\scriptstyle #1$})}} \kern-1pt}
\def\del{\nabla}
\def\div{\mathop{\rm div}\nolimits}    
\def\Scurl{\mathop{\rm Scurl}\nolimits} 

\begin{document}

\title[Relative-observer definition of the Simon tensor]
{Relative-observer definition of the Simon tensor}

\author{
Donato Bini${}^{\dag}$ and
Andrea Geralico${}^{\dag}$
}
\address{
  ${}^{\dag}$\
 Istituto per le Applicazioni del Calcolo \lq\lq M. Picone" via dei Taurini 19 I-00185 Rome (IT)
}

\begin{abstract}
The definition of Simon tensor, originally given only in the Kerr spacetime and associated with the static family of observers, is generalized to any spacetime and to any possible observer family.
Such generalization is obtained by a standard \lq\lq 3+1" splitting of the Bianchi identities, which are rewritten here as a  \lq\lq balance equation" between various spatial fields, associated with the kinematical properties of the observer congruence and representing the spacetime curvature. 
\end{abstract}

\pacno{04.20.Cv}

\section{Introduction}

In 1984 Simon \cite{Simon} provided an invariant characterization of Kerr spacetime in terms of the vanishing of the so-called \lq\lq Simon tensor." 
The latter is associated with both the temporal Killing properties of the spacetime and its Petrov type D character, in the sense that its vanishing is an equivalent formulation of the fact that 
the principal null directions of the covariant derivative of the temporal Killing vector (i.e., the associated \lq\lq Papapetrou field") are exactly the same of the Kerr spacetime itself \cite{krisch88,Mars99,Fasop99,Mars00,Fasop00,Mars01}.
The underlying geometrical structure relating these two apparently separated concepts, that is the existence of a temporal Killing vector and the spectral type, is the $3+1$ form of the Bianchi identities or, more in general, the $3+1$ form of the spacetime divergence equation for the Weyl tensor, explicitly evaluated in the Kerr spacetime by using the family of static observers \cite{Bini:2001ke,Bini:2004qf}.
In fact, the $3+1$ form of the Bianchi identities when the observer family considered for the splitting are the static observers (whose four-velocity is aligned with the temporal Killing congruence) implies that the Simon tensor (which is of differential type, like a curl) coincides with what is often referred to as the Simon-Mars tensor (which is of algebraic type, like a vector product), according to a terminology which we will introduce below. 
The divergence equations for the Weyl tensor state that the curl is equal to the vector product term and the Kerr spacetime is so special that, besides being equal in general,  these two terms are both identically vanishing.

In the literature the existence of similar properties related to both the Simon and Simon-Mars tensors has been explored in other spacetimes with Killing symmetries, even nonvacuum \cite{Kramer85,Mars:2001ge,Ferrando:2010bd,Some:2014kfa,Mars:2016vtr,Mars:2016ahh,Mars:2016ynw,Paetz:2017lkn}.
These tensors have been used there to provide a characterization of the considered spacetimes.
For instance, the Simon-Mars tensor has been shown in Ref. \cite{Some:2014kfa} to measure (locally) the deviation of a given stationary spacetime from the Kerr one, including general spacetimes which have a matter content and numerically generated ones.
Furthermore, the vanishing of the Simon-Mars tensor has been used in Ref. \cite{Mars:2016vtr} and related works to investigate the uniqueness and stability properties of spacetimes having the same asymptotic structure of a Kerr-de Sitter solution and to classify them.
What has not been considered yet is instead the role of different families of observers even in the same context of the Kerr spacetime, i.e., a relative-observer definition of the Simon and Simon-Mars tensors. The divergence equation for the Weyl tensor is indeed richer within this perspective, and can be rewritten as a \lq\lq balance equation" between an observer-dependent version of both the Simon and Simon-Mars tensors in terms of spatial tensors  associated with the kinematical properties of the observer congruence, namely acceleration, vorticity and expansion. 
This balance equation allows for an additional characterization of (known) special families of timelike congruences in the Kerr spacetime, besides the static observers. Here we will explore the cases of the Zero Angular Momentum Observers (ZAMOs), the Carter family of observers and the Painlev\'e-Gullstrand (PG) observers.
We will repeat then this analysis also for the Kerr-de Sitter spacetime as a simple example of a non-vacuum solution.

We use geometrical units and follow the notation and convention of \cite{Jantzen:1992rg} for the spacetime splitting of tensor fields. To make the paper self-contained the main basic relations of $3+1$ splitting are shortly recalled too.

\section{Observer congruences and associated $3+1$ spacetime splitting}

Consider a spacetime admitting a congruence of timelike world lines with unit tangent vector $u$ ($u_\alpha u^\alpha =-1$) associated with a given observer family.
Its splitting into space-plus-time is accomplished by the orthogonal decomposition of the tangent space at any point into a local time direction (along $u$) and an orthogonal 3-space (the local rest space LRS$_u$).
This can be obtained by the systematic application of temporal ($T(u)$) and spatial ($P(u)$) projections defined by 
\beq
T(u)^\alpha{}_\beta=-u^\alpha u_\beta\,,\qquad
P(u)^\alpha {}_\beta=\delta^\alpha{}_{\beta}+u^\alpha u_\beta\,,
\eeq
which are related to the metric tensor as
\beq
g_{\alpha\beta} =  T(u)_{\alpha\beta} + P(u)_{\alpha\beta}\,.
\eeq
Tensors with no components along $u$ are called spatial with respect to $u$. 
The spatial metric $P(u)_{\alpha\beta}$ corresponds to a Riemannian metric on the observer LRS$_u$. 

The congruence $u$ is itself characterized by the kinematical quantities (acceleration $a(u)$, expansion $\theta(u)$, vorticity $\omega(u)$) 
which result from the splitting of its covariant derivative $\nabla u=(\nabla u)_{\alpha\beta}=\nabla_\beta u_\alpha=u_{\alpha;\beta}$, i.e.,
\begin{eqnarray}
 a (u)_\alpha &=& u_{\alpha\,;\,\beta}\,u^\beta \ ,\nonumber \\
 \theta (u)_{\alpha\beta} &=& P(u)^\gamma {}_\alpha\,
  P(u)^\delta {}_\beta\,u_{\,(\,\gamma\,;\,\delta\,)}\ ,\nonumber \\
 \omega (u)_{\alpha\beta} &=&- P(u)^\gamma {}_\alpha\,
  P(u)^\delta {}_\beta\,u_{\,[\,\gamma\,;\,\delta\,]} \ ,
\end{eqnarray}
with ${\rm Tr}\, \theta(u)= \Theta (u)$.
 The unit volume 3-form 
$\eta (u)_{\alpha\beta\gamma} 
= u^\delta \,\eta_{\delta\alpha\beta\gamma}$ (with $\eta_{\delta\alpha\beta\gamma}=\sqrt{-g}\epsilon_{\delta\alpha\beta\gamma}$, $\epsilon_{0123}=1$)
can be used to define a spatial duality
operation for antisymmetric spatial tensor fields
\beq
\dualp{u} A_\alpha =\frac12 \eta(u)_{\alpha\beta\gamma}A^{\beta\gamma}\,,\qquad
\dualp{u} A_{\alpha\beta} =\eta(u)_{\alpha\beta\gamma}A^\gamma\ .
\eeq
The vorticity vector field $\omega(u)^\alpha=\frac12\eta(u)^{\alpha\beta\gamma}\omega(u)_{\beta\gamma}$ is defined as the spatial dual of the vorticity 2-form. In index-free notation these will be denoted respectively by $\omega(u)$ and $\dualp{u}\omega(u)$. 

One can also define a spatial cross product for a spatial vector ($X$) and a symmetric spatial tensor ($A$) and for two symmetric spatial tensor fields
($A, B$)
\begin{eqnarray}
  [X \times_u A]^{\alpha\beta} &=& \eta (u)^{\,\gamma\delta (\alpha }
  X_{\gamma} A^{\beta)} {}_{\delta}\ ,\nonumber\\
{} [A\times_u B]_{\alpha} &=& \eta (u)_{\alpha\beta\gamma}
   A^{\beta} {}_{\delta}\, B^{\delta\gamma}\ .
\end{eqnarray}
In a similar way one can introduce a spatial inner product  for a spatial vector ($X$) and a symmetric spatial tensor ($A$) and for two spatial symmetric tensors ($A, \, B$)
\begin{eqnarray}
[X\cdot A]_\alpha &=&P(u)_{\beta\mu}X^\beta A^\mu{}_\alpha 
= X_\mu A^\mu{}_\alpha\ , \nonumber\\
{} [A\cdot B]_{\alpha\beta} 
&=& A_{(\alpha|\mu}P(u)^{\mu\nu} B_{\nu|\beta)}
= A_{(\alpha|\mu}B^{\mu}{}_{\beta)}\ .
\end{eqnarray}
Spatial projection of differential operators also leads to the following \lq\lq spatial differential operators'' \cite{Jantzen:1992rg}: the spatial covariant derivative
\beq
\del (u)_\alpha X_\beta =P(u)^\mu_\alpha P(u)^\nu_\beta \del_\mu X_\nu \ ,
\eeq
which in turn allows the introduction of the
generalized divergence  and (symmetrized) curl operations for spatial symmetric tensor fields by
\begin{eqnarray}
\label{eq:scurl}
  [\div (u) A]^\alpha 
&=& \del(u)^\beta A^\alpha {}_\beta \ ,\nonumber\\
{}[\Scurl (u) A]^{\alpha\beta}
  &=&\eta(u)^{\gamma\delta\,(\,\alpha} \del(u)_\gamma A^{\beta\,)} {}_\delta \ ,
\end{eqnarray}
and the spatial Lie derivative
\beq
\del_{\rm(lie)} (u) X_\alpha = P(u)_\alpha^\beta \pounds_u X_\beta\,.
\eeq
These may be extended to any rank tensors in a standard way. 
The noncommutativity of the Lie derivative with raising and lowering indices requires when using abstract notation to specify the index position of any tensor. We use the notation ${}^\flat$ and ${}^\sharp$ to denote the completely covariant and completely contravariant form of a tensor, respectively.

Both the vector-tensor cross product and the Scurl operator annihilate the pure trace part of the symmetric spatial tensor
\begin{eqnarray}
&& 
 X \times_u A = X \times_u A^{\rm(TF)}\ ,\quad
 \Scurl A = \Scurl A^{\rm(TF)}\ .
\end{eqnarray}
Furthermore, one can show that a vanishing cross product $X \times_u A^{\rm(TF)}=0$ aligns $X$ and $A^{\rm(TF)}$ \cite{Bini:2001ke}, i.e.,
\beq\label{eq:align}
  A^{\rm(TF)} \propto [X\otimes X]^{\rm(TF)}\ .
\eeq
Here TF denotes the trace-free part of a tensor, i.e., for a spatial tensor 
\beq
A^{\rm(TF)}_{\alpha\beta} 
= A_{\alpha\beta} -{\textstyle\frac13} P(u)_{\alpha\beta} A^\gamma{}_\gamma\,,
\eeq
and similarly STF the trace-free part of a symmetrized tensor with respect to all its indices.

\section{The $3+1$ form of divergence equation for the Weyl tensor}

The Weyl tensor $C_{\alpha\beta\gamma\delta}$
is the part of the spacetime curvature which is not directly determined by the energy-momentum tensor.
Using its tracefree property, the once-contracted Bianchi identities determine the divergence of the Weyl tensor \cite{Kramer}
\beq
0= 3 R^{\alpha\beta}{}_{[\gamma\delta;\epsilon]} \delta^\gamma{}_\alpha
     = R^{\alpha\beta}{}_{\delta\epsilon;\alpha}
           + 2 R^\beta{}_{[\delta;\epsilon]} 
     = C^{\alpha\beta}{}_{\delta\epsilon;\alpha}
           + \frac12\, R^\beta{}_{\delta\epsilon} \ ,
\eeq
where
\beq
     R^\beta{}_{\delta\epsilon}
         = 2 \left( R^\beta{}_{[\delta}
             -\frac16R \delta^\beta{}_{[\delta} \right){}_{;\epsilon]}
\eeq
is the (spacetime) Cotton tensor.
Half the Cotton tensor 
is a $1\choose2$-tensor (like a current, in analogy with electromagnetism, where $F^{\alpha\beta}{}_{;\beta} = 4\pi J^\alpha$)
\beq
\label{current}
 J^\alpha{}_{\beta\gamma} =C^{\alpha\delta}{}_{\beta\gamma;\delta} =
                           -\del_{[\beta} \left( R^\alpha{}_{\gamma ]}
                           -\frac16 R \delta^\alpha{}_{\gamma ]} \right) \ ,
\eeq
which vanishes identically in vacuum.
Splitting these equations with respect to a generic
congruence $u$ leads to their $3+1$ relative-observer form  \cite{Ellis99}.

Splitting the Weyl tensor yields the two symmetric tracefree spatial fields
\begin{eqnarray}
  E(u)^{\alpha} {}_{\beta} 
&=&  C^{\alpha} {}_{\gamma\beta\delta}\, u^\gamma \, u^\delta \ ,\nonumber\\
  H(u)^{\alpha} {}_{\beta} 
&=& - {}^\ast C^{\alpha} {}_{\gamma\beta\delta}\, u^\gamma \, u^\delta
= \frac{1}{2}\, \eta \, ^{\alpha} {}_{\gamma} {}^{\delta}\,
 C^{\gamma}{}_{\delta\beta\rho}\, u^\rho 
\ ,
\end{eqnarray}
which are called its electric and magnetic parts, respectively, and are used to classify the gravitational field by Petrov type.
Hereafter we will drop the dependence on $u$ of all spatial fields introduced above to ease notation.
Splitting Eq. (\ref{current}) and using adapted frame component notation gives (see Eq. (3.7) in Ref. \cite{Bini:2001ke})
\beq\eqalign{
   J^\top{}_{a\top}
&=[\div E]_{a}+3 [\omega \cdot H]_a
+ [\theta \times H]_a 
\ ,\cr
   J^{*} {}^\top{}_{a\top}
&=-[\div H]_{a} + 3 [\omega \cdot E]_a
+[\theta \times E]_a
\ ,\cr
   J_{(\,ab\,)\,\top} 
&= [\Scurl H +2a\times H]_{ab} - \del_{\rm(lie)} (u) E_{ab}
      -[\omega\times E]_{ab} 
\cr
      &- 2\Theta E_{ab} +5[\theta\cdot E]_{ab} -P(u)_{ab}{\mathrm{Tr}}[\theta\cdot E]
\ ,\cr
   J^{*} {}_{(\,ab\,)\,\top} 
&= [\Scurl E +2a\times E]_{ab} + \del_{\rm(lie)} (u) H_{ab}
       +[\omega\times H]_{ab} 
\cr
      &+ 2\Theta H_{ab} -5[\theta\cdot H]_{ab} 
  + P(u)_{ab}{\mathrm{Tr}}[\theta\cdot H]
\ ,\label{MaxLike1}
}\eeq
with $J^\top{}_{\beta \top}\equiv J^u{}_{\beta u}=J^\mu{}_{\beta \nu}u_\mu u^\nu$, etc.
In terms of $Z=E+iH$ and $z=-a+i\omega$ the above formulas become\footnote{
In the literature the divergence equation of the Weyl tensor is written either in terms of the  fields $Z$ and $z$ or in terms of their complex conjugate. This is mainly due to the sign conventions entering the definition of both the vorticity vector and the magnetic part of the Weyl tensor. In order to make the present paper self-contained we have included  the  definition of all used quantities.}
\begin{eqnarray}
\label{MaxLike1fin}
\rho^{\rm(G)}&=&\div Z+3i\omega\cdot Z+i\theta\times Z\,,\\
\label{MaxLike2fin}
J^{\rm(G)}&=&\Scurl Z+a\times Z-z\times Z-i(\del_{\rm(lie)} (u) Z+\Theta Z)\nonumber\\
&&
-i[\Theta Z+P(u){\rm Tr}(\theta\cdot Z)]+5i\theta\cdot Z\,,
\end{eqnarray}
where
\begin{eqnarray}
\rho_\alpha^{\rm(G)}(u)&=&P(u)^\beta{}_{\alpha}J^u{}_{\beta u}+iJ^{*} {}^u{}_{\beta u}\,,\nonumber\\
J_{\alpha\beta}^{\rm(G)}(u)&=&P(u)^\mu{}_\alpha P(u)^\nu{}_\beta[ J^{*} {}_{(\,\mu\nu\,)\,u}-iJ_{(\,\mu\nu\,)\,u}]\,.
\end{eqnarray}

We will focus below on the \lq\lq Scurl'' equation (\ref{MaxLike2fin}) only, which can be written in compact form as
\beq
\label{eqbal}
J^{\rm(G)}=-i\del_{\rm(lie)} (u) Z+S+S_a+S_z+S_\theta\,,
\eeq
by introducing the following notation
\begin{eqnarray}
S&=&\Scurl Z\,,\qquad
S_{a}=a\times Z\,,\qquad
S_{\omega}=\omega\times Z\,,\nonumber\\
S_\theta&=&-i[2\Theta Z+P(u){\rm Tr}(\theta\cdot Z)-5\theta\cdot Z]\,,
\end{eqnarray}
with $S_z=-z\times Z=S_{a}-iS_{\omega}$.
Furthermore, we have that $\del_{\rm(lie)} (u) P(u)^\sharp=-2\theta(u)^\sharp$, implying that
\beq
{\rm Tr}(\del_{\rm(lie)} (u) Z)=2{\rm Tr}(\theta\cdot Z)\,,
\eeq
whence
\beq
{\rm Tr}[-i\del_{\rm(lie)} (u) Z+S_\theta]=0\,.
\eeq
Eq. (\ref{eqbal}) in vacuum ($\rho^{\rm(G)}=0$, $J^{\rm(G)}=0$) reduces to a balance equation between the various terms.
It is clear that in this case a change of observer simply implies a re-shuffling of the individual contributions. 
On the other hand, it is convenient to have only trace-free terms in the balance equation by combining the Lie-derivative term and the expansion term in a single  trace-free term, $S_{\rm (lie)\theta}$, namely
\beq
S_{\rm (lie)\theta}=-i\del_{\rm(lie)} (u) Z+S_\theta\,,
\eeq
so that
\beq
\label{eqbalance}
S+S_a+S_z+S_{\rm (lie)\theta}=0\,,
\eeq
where it is understood that $S=S(u)$, $S_a=S_a(u)$, etc., since all these trace-free spatial tensors refer to a generic observer $u$ in a completely general spacetime.
Special observers then arise associated with the vanishing of each term. 
The most natural families are:
  
\begin{enumerate}

  \item {\it Geodesic observers} ($S_a=0$, so that $S_z=-iS_\omega$):
\begin{eqnarray}\fl\qquad
0&=&\div Z+3i\omega\cdot Z+i\theta\times Z\,,\nonumber\\
\fl\qquad
0&=&S-iS_\omega+S_{\rm (lie)\theta}\,,
\end{eqnarray} 
  
\item  {\it Vorticity-free observers} ($S_\omega=0$, so that $S_z=S_a$):
\begin{eqnarray}\fl\qquad
0&=&\div Z+i\theta\times Z\,,\nonumber\\
\fl\qquad
0&=&S+2S_a+S_{\rm (lie)\theta}\,,
\end{eqnarray}

  \item {\it Expansion-free observers} ($S_\theta=0$, so that $S_{\rm (lie)\theta}=-i\del_{\rm(lie)} (u) Z$):
 \begin{eqnarray}\fl\qquad
0&=&\div Z+3i\omega\cdot Z\,,\nonumber\\
\fl\qquad
0&=&S+S_a+S_z-i\del_{\rm(lie)} (u) Z\,,
\end{eqnarray}

\item {\it Geodesic and irrotational observers} ($S_a=0=S_\omega$, so that $S_z=0$ too):
\begin{eqnarray}\fl\qquad
0&=&\div Z+i\theta\times Z\,,\nonumber\\
\fl\qquad
0&=&S+S_{\rm (lie)\theta}\,.
\end{eqnarray} 

\end{enumerate} 

In the non-vacuum case the decomposition (\ref{eqbal}) still holds, but in this case $J^{\rm(G)}\not=0$ in general.
One can use the Einstein equations to replace the Ricci tensor and scalar curvature terms by the energy-momentum tensor in Eq. (\ref{current}), which becomes
\beq
\label{JTmunu}
 J^\alpha{}_{\beta\gamma} =-\kappa\del_{[\beta} \left( T^\alpha{}_{\gamma ]}-\frac13 T \delta^\alpha{}_{\gamma ]} \right) \,.
\eeq

Let us complete this section recalling the expression of $Z(u)$ in terms of the kinematical fields of the observer, namely
\beq
Z(u)=Z_z(u)+Z_\theta(u)\,,
\eeq
where
\begin{eqnarray}
Z_z(u)&=&[-\nabla(u)z(u)+z(u)\otimes z(u)]^{\rm STF}
\,,\nonumber\\
Z_\theta(u)&=&-[\del_{\rm(lie)} (u) \theta(u)^\flat]^{\rm STF}+[\theta(u)^2]^{\rm STF}-2[\theta(u)\cdot\omega(u)]^{\rm STF}\nonumber\\
&&
+i\Scurl \theta(u)\,.
\end{eqnarray}
Noticeably, expansion-free observers have $Z_\theta(u)=0$ and hence $Z(u)=Z_z(u)$.
As it is known static observers in Kerr spacetime are so special that besides forming an expansion-free congruence of world lines have the additional property that $[\nabla(u)z(u)]^{\rm STF}\propto [z(u)\otimes z(u)]^{\rm STF}$, implying in turn the simple relation $Z(u)\propto[z(u)\otimes z(u)]^{\rm STF}$ (see below).

\section{Special observes and adapted frames in the Kerr spacetime}

Let us consider the Kerr spacetime, whose metric written in standard Boyer-Lindquist coordinates $(t,r,\theta,\phi)$ is given by
\begin{eqnarray}
ds^2 &=& -\left(1-\frac{2Mr}{\Sigma}\right)dt^2 
-\frac{4aMr}{\Sigma}\sin^2\theta dtd\phi+ \frac{\Sigma}{\Delta}dr^2\nonumber\\
&&+\Sigma d\theta^2+\frac{A}{\Sigma}\sin^2 \theta d\phi^2\,,
\end{eqnarray}
with $\Delta=r^2-2Mr+a^2$, $\Sigma=r^2+a^2\cos^2\theta$ and $A=(r^2+a^2)^2-\Delta a^2\sin^2\theta$.
Here $a$ and $M$ denote the specific angular momentum and the total mass of the spacetime solution.
The inner and outer horizons are located at $r_\pm=M\pm\sqrt{M^2-a^2}$.

There exist at least three natural observer families associated with the Kerr geometry, which can be easily described in terms of the Boyer-Lindquist coordinates because they are adapted to the Killing symmetries. The static observers follow the integral curves of the (stationary) Killing vector field $\partial_t$, while the world lines of the ZAMOs are orthogonal to the time coordinate hypersurfaces. Finally, the Carter observers play a key role to the separability of the geodesic equations and are of fundamental importance for the algebraic properties of the curvature tensor \cite{carter}. All three families differ only by relative azimuthal motion, and hence their natural adapted frames are all related by relative boosts in the $t$-$\phi$ plane of the tangent space.

We will also consider a fourth family of observers, the Painlev\'e-Gullstrand family \cite{Painleve21,Gullstrand22,doran,cook,hamilton,Bini:2014uua,Bini:2015wpa}, who move radially with respect to the ZAMOs and form a geodesic and irrotational congruence. Their adapted coordinates have the very useful property of remaining valid inside the outer horizon,
leading to the terminology of \lq\lq horizon-penetrating coordinates.''
Furthermore, in the Schwarzschild case the intrinsic geometry of the associated time slices is flat. 

The kinematical properties of all these observers as well as the components of the electric and magnetic parts of the Weyl tensor are reviewed in Appendix A.

\subsection{Static observers}

The static observers, which exist only in the spacetime region outside the ergosphere where $g_{tt}<0$,  form a congruence of accelerated, nonexpanding and locally rotating world lines. 
They have 4-velocity $u=m$, where
\beq
\label{thd}
m
=\left(1-\frac{2Mr}{\Sigma}\right)^{-1/2}\,\partial_t
\equiv L_m^{-1}\,\partial_t\,.
\eeq
An orthonormal frame adapted to $m$ is
\begin{eqnarray}
\label{static_triad}
e(m)_1 &=&
\sqrt{\frac{\Delta}{\Sigma}} \,\partial_r
\equiv e_{\hat r}
\,,\qquad
e(m)_2 = 
\frac{1}{\sqrt{\Sigma}}\,\partial_\theta
\equiv e_{\hat \theta}
\,,\nonumber\\
e(m)_3 &=&
\frac{\sqrt{\Sigma-2Mr}}{\sin \theta \sqrt{\Delta \Sigma}} \left(\partial_\phi-\frac{2Mar\sin^2\theta}{\Sigma-2Mr} \partial_t  \right)
\,. \nonumber\\
\end{eqnarray}

\subsection{ZAMOs}

The ZAMOs are instead locally nonrotating and exist everywhere outside of the outer horizon.
They have 4-velocity $u=n$, where
\begin{eqnarray}\label{ncoord}
n
&=& \sqrt{\frac{A}{\Delta\Sigma}}\,\left(\partial_t+\frac{2aMr}{A}\partial_\phi\right)
\,.
\end{eqnarray}
The normalized spatial coordinate frame vectors 
\begin{eqnarray}\label{zamotriad}
e(n)_1&=&e_{\hat r}\,, \qquad
e(n)_2=e_{\hat \theta}\,, \qquad
e(n)_3=
\frac{\sqrt{\Sigma}}{\sin\theta \sqrt{A}} \,\partial_\phi
\equiv e_{\hat \phi}
\end{eqnarray}
together with $n$ form an orthonormal adapted frame.

\subsection{Carter observers}

Carter observers have $4$-velocity $u=u_{\rm (car)}$ given by
\begin{eqnarray}
\label{car_obs}
u_{\rm (car)} &=&\frac{r^2+a^2}{\sqrt{\Delta \Sigma}}\left(\partial_t +\frac{a}{r^2+a^2}\,\partial_\phi  \right)
\,.
\end{eqnarray}
A spherical orthonormal frame adapted to $u_{\rm (car)}$ is obtained by using the triad boosted from the either the ZAMO or static observer spherical frame along the azimuthal direction, so that
\beq\fl\quad
e(u_{\rm (car)})_1 = e_{\hat r}\,,\quad 
e(u_{\rm (car)})_2 = e_{\hat \theta}\,,\quad
e(u_{\rm (car)})_3  = \frac{a\sin \theta}{\sqrt{\Sigma}}\left(\partial_t +\frac{1}{a\sin^2\theta}\,\partial_\phi  \right)\,. 
\eeq

\subsection{Painlev\'e-Gullstrand observers}

In the Kerr spacetime it is also interesting to study the PG geodesic and irrotational family of orbits.
The associated  four velocity 1-form, denoted by $u_{\rm(PG)}^\flat$, is given by
\beq
\label{PG4vel}
u_{\rm(PG)}^\flat=-dt -\frac{\sqrt{2Mr(r^2+a^2)}}{\Delta}\, dr\ . 
\eeq
It is easy to see that
\beq
\label{calN}
u_{\rm(PG)}=\gamma{(u_{\rm(PG)},n)}[n+\nu(u_{\rm(PG)},n) e_{\hat r}]\ ,
\eeq
so that the PG observers move radially with respect to the ZAMOs with a 
relative speed 
\beq
\nu(u_{\rm(PG)},n)=-\sqrt{\frac{2Mr(r^2+a^2)}{A}}\ , \quad
\gamma{(u_{\rm(PG)},n)}=\sqrt{\frac{A}{\Delta\Sigma}}\ .
\eeq
A  frame adapted to the PG observers can be fixed with the triad
\begin{eqnarray}
\label{framePG}
e(u_{\rm(PG)})_1&=&\gamma{(u_{\rm(PG)},n)}[\nu(u_{\rm(PG)},n) n+e(n)_1]\ , \nonumber\\
e(u_{\rm(PG)})_2&=& e(n)_2\ , \qquad
e(u_{\rm(PG)})_3= e(n)_3\ .
\end{eqnarray}

\section{The relative-observer definition of the Simon tensor}

The original definition of both Simon and Simon-Mars tensors dealt with expansion-free observers in Kerr spacetime, i.e., $\theta(m)=0$ (which implies $S_\theta(m)=0$). The main equation (\ref{eqbalance}) thus becomes
\beq
S(m)+S_a(m)+S_z(m)=0\,,
\eeq
since $\del_{\rm(lie)} (m) Z=0$ for a static congruence of observers (so that $S_{\rm(lie)\theta}(m)=0$).
Moreover, $m = L_m^{-1}\xi$ with the associated Killing vector $\xi=\partial_t$ (see Eq. (\ref{thd})), so that
the acceleration can be expressed as $a(m)= \nabla(m)\ln L_m$.
Incorporating the scale factor $L_m$ into the Scurl leads to 
\beq\fl\qquad
L_m^{-1}{\rm Scurl}[L_mZ(m)]=S(m)+S_a(m)=-S_z(m)=z(m)\times Z(m)\,,
\eeq
where the lhs term is identified with the Simon tensor, whereas the rhs tensor with the Simon-Mars tensor.
Their equality allows to evaluate one in terms of the other. In other words, while the Simon tensor concerns the differential properties of the curvature, the Simon-Mars tensor connects it with the algebraic properties of the curvature.
The most important fact is that these two tensors both vanish.
In fact, it turns out that $S_a(m)=-S(m)$ implying that the Simon tensor vanishes and $S_\omega(m)=iS(m)$, so that the Simon-Mars tensor $S_z(m)=0$ too.
The geometrical meaning of such a vanishing has been understood in terms of alignment of the principal null directions of either $Z(m)$ and $z(m)$ \cite{Fasop99,Fasop00,Bini:2001ke}.
The nonvanishing frame components of $S(m)$ are given by
\beq\fl\qquad
S(m)_{13}=-\frac32aM^2\Delta\sin\theta\frac{ir-a\cos\theta}{(\Sigma-2Mr)^2\Sigma^{5/2}}\,,\quad
S(m)_{23}=\frac{ia\sin\theta}{\sqrt{\Delta}}S(m)_{13}\,.
\eeq

The balance equation (\ref{eqbalance}) is completely general for any spacetime and any observer family $u$, and can be read in different ways.
It is convenient to keep $S_z(u)$  as the \lq\lq generalized Simon-Mars tensor'' 
\beq
\label{SMtens}
S_{\rm Simon-Mars}(u)=S_z(u)\,,
\eeq
which retains (for $z(u)\not=0$) the simple geometrical meaning of alignment of the principal null directions of both $Z(u)$ and $z(u)$.
As a consequence, Eq. (\ref{eqbalance}) also defines a \lq\lq generalized Simon tensor'' as 
\beq
\label{simontens}
S_{\rm Simon}(u)=S(u)+S_a(u)+S_{\rm (lie)\theta}(u)\,,
\eeq
the Bianchi identities providing the vanishing of their sum
\beq
\label{equivrel}
S_{\rm Simon}(u)+S_{\rm Simon-Mars}(u)=0\,,
\eeq
which is the observer-dependent content of the Bianchi identities in any spacetime. 
Different observers will measure different Simon and Simon-Mars tensors, but their sum is observer-independent and always identically zero.
Note that the Simon-Mars tensor $S_z(u)$ vanishes when the two principal null directions of the (complex) 2-forms 
\beq
{\mathcal P}(u)=u\wedge z(u)+{}^{*(u)}z(u)
\eeq
are aligned with those of $Z(u)$.
${\mathcal P}(u)$ results from 
\beq\fl\quad
{\mathcal P}(u)={\mathcal F}(u)-i{}^{*(u)}{\mathcal F}(u)\,,\quad
{\mathcal F}(u)=\frac12[du-u\wedge a(u)]=-u\wedge a(u)+{}^{*(u)}\omega(u)\,.
\eeq
In the case of a static observer in Kerr spacetime ${\mathcal F}(m)$ becomes proportional to the Papapetrou field $\xi_{[\mu;\nu]}$ and allows for the well known interpretation \cite{Mars99,Fasop99}.

The equivalence (\ref{equivrel}) between the Simon and Simon-Mars tensors is broken in the nonvacuum case due to the presence of energy-momentum terms, implying that the Cotton gravitational current is nonzero in general. However, one can restore the equality by absorbing these source terms into a redefinition of the Simon tensor (\ref{simontens}), i.e., by including the additional spatial field $-J^{\rm(G)}(u)$ in the rhs. This is the case of electrovac stationary spacetimes which contain a source-free electromagnetic field as well as stationary spacetimes generated by rigidly rotating perfect fluids discussed in Ref. \cite{Bini:2004sg}. In contrast, when the matter-energy content is given by a cosmological constant term the associated Cotton current is identically zero for any observer family. In fact, the energy-momentum tensor $T_{\mu\nu}=\Lambda g_{\mu\nu}$ leads to a vanishing tensor $J^\alpha{}_{\beta\gamma}$ (see Eq. (\ref{JTmunu})). $\Lambda$ will clearly enter all the splitting fields once a specific solution of the Einstein's field equations is considered. Appendix B contains an explicit example.

We will consider below the decomposition of the Simon and Simon-Mars tensors in the Kerr spacetime with respect to the various families of observers introduced above.

\subsection{ZAMOs}

We have $\omega(n)=0$, $\Theta(n)=0$ and ${\rm Tr}(\theta(n)\cdot Z(n))=0$, so that $S_\omega(n)=0$ and $S_\theta(n)=5i\theta(n)\cdot Z(n)$, and
\beq
S(n)+2S_a(n)+S_{\rm (lie)\theta}(n)=0\,.
\eeq
In this case 
\beq\fl\qquad
S_{\rm Simon}(n)=S(n)+S_a(n)+S_{\rm (lie)\theta}(n)\,,\qquad
S_{\rm Simon-Mars}(n)=-S_a(n)\,.
\eeq
The nonvanishing components of $S(n)$, $S_a(n)$ and $S_{\rm (lie)\theta}(n)$ are listed below
\begin{eqnarray}\fl\qquad
\label{Sazamo}
[S_a(n)]_{13}&=&-\frac{3(r^2+a^2)aM^2\sin\theta}{2A^2\Sigma^{3/2}(ir+a\cos\theta)^3}\{
-2iar(r^2+a^2)^2\cos\theta
\nonumber\\
\fl\qquad
&&
+2r^2(r^2+a^2)\Delta-[(r^2+a^2)^2-4Mr^3]\Sigma\}
\,,\nonumber\\
\fl\qquad
{}[S_a(n)]_{23}&=&\frac{ia\sqrt{\Delta}\sin\theta}{r^2+a^2}[S_a(n)]_{13}\,,
\end{eqnarray}
and
\begin{eqnarray}\fl\qquad
[S_{\rm(lie)\theta}(n)]_{13}&=&-\frac{3aM^2\sin\theta}{2A^2\Sigma^{3/2}(ir+a\cos\theta)^3}\{
10ir(r^2+a^2)\Delta a^3\sin^2\theta\cos\theta\nonumber\\
\fl\qquad
&&
-[(r^2-a^2)\Sigma+2r^2(r^2+a^2)][3(r^2+a^2)^2
+2\Delta a^2\sin^2\theta]
\}
\,,\nonumber\\
\fl\qquad
{}[S_{\rm(lie)\theta}(n)]_{23}&=&\frac{3a^2M^2\sqrt{\Delta}\sin^2\theta}{2A^2\Sigma^{3/2}(ir+a\cos\theta)^3}\{
5i(r^2+a^2)[(r^2-a^2)\Sigma+2r^2(r^2+a^2)]\nonumber\\
\fl\qquad
&&
-2ar\cos\theta[3\Delta\Sigma-(r^2+a^2)(5(r^2+a^2)-6Mr)]
\,,
\end{eqnarray}
and
\begin{eqnarray}\fl\qquad
S(n)_{13}&=&-\frac{3aM^2\sin\theta}{2A^2\Sigma^{3/2}(ir+a\cos\theta)^3}\{
6r^2(r^2+a^2)^3\nonumber\\
\fl\qquad
&&
-(r^2-a^2)\Sigma[2\Delta\Sigma-(r^2+a^2)(3(r^2+a^2)-4Mr)]\nonumber\\
\fl\qquad
&&
+2iar(r^2+a^2)\cos\theta[5\Delta\Sigma-(r^2+a^2)(3(r^2+a^2)-10Mr)]
\}
\,,\nonumber\\
\fl\qquad
S(n)_{23}&=&\frac{3ia^2M^2\sqrt{\Delta}\sin^2\theta}{2A^2\Sigma^{3/2}(ir+a\cos\theta)^3}\{
6ir\Delta a^3\sin^2\theta\cos\theta\nonumber\\
\fl\qquad
&&
-[(r^2+a^2)(7r^2-3a^2)-8Mr^3]\Sigma-2r^2(r^2+a^2)[3(r^2+a^2)+4Mr]
\}\,.\nonumber\\
\fl\qquad
\end{eqnarray}

\begin{figure}
\centering
\includegraphics[scale=0.35]{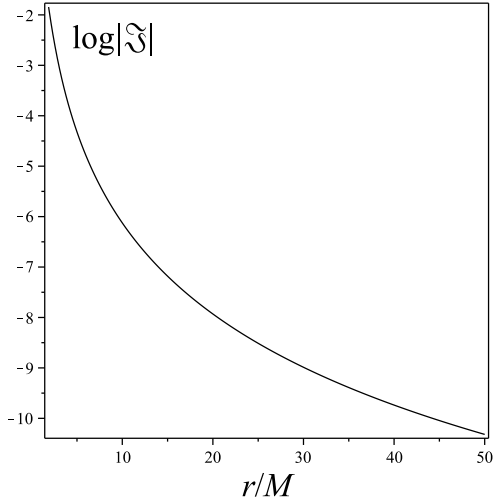}
\caption{\label{fig:1} 
The behavior of the (log10 of) the absolute value of the imaginary part $\mathcal{J}$ of the frame component $[S_a(n)]_{13}$ is shown as a function of the radial coordinate for $\theta=\pi/2$ and $a/M=0.5$.
It never vanish, implying that the Simon-Mars tensor is always different from zero as measured by ZAMOs. 
}
\end{figure}

The Simon and Simon-Mars tensors both vanish if $[S_a(n)]_{13}=0$. Eq. (\ref{Sazamo}) implies that its real part identically vanishes for $\theta=\pi/2$, whereas its imaginary part is always different from zero, as shown in Fig. \ref{fig:1}.


\subsection{Carter observers}

We find $S_\omega(u_{\rm(car)})=0$ (but $\omega(u_{\rm(car)})\not=0$) and $\Theta(u_{\rm(car)})=0={\rm Tr}(\theta(u_{\rm(car)})\cdot Z(u_{\rm(car)}))=0$, so that $S_\theta(u_{\rm(car)})=5i\theta(u_{\rm(car)})\cdot Z(u_{\rm(car)})$, and
\beq
S(u_{\rm(car)})+2S_a(u_{\rm(car)})+S_{\rm (lie)\theta}(u_{\rm(car)})=0\,,
\eeq
with
\begin{eqnarray}
S(u_{\rm(car)})&=&\frac32\frac{iMa\sqrt{\Delta}\sin\theta(3ir-2a\cos\theta)}{\Sigma^2(ir+a\cos\theta)}
[e(u_{\rm (car)})_1\otimes e(u_{\rm (car)})_3\nonumber\\
&&
+e(u_{\rm (car)})_3\otimes e(u_{\rm (car)})_1]\,,
\end{eqnarray}
and
\begin{eqnarray}
S_a(u_{\rm(car)})&=&\frac{a\cos\theta}{3ir-2a\cos\theta}S(u_{\rm(car)})\,,\nonumber\\
S_{\rm(lie)\theta}(u_{\rm(car)})&=&-\frac{3ir}{3ir-2a\cos\theta}S(u_{\rm(car)})\,.
\end{eqnarray}
In this case these three tensor are all proportional (just as the electric and magnetic parts of the Weyl tensor) and
\begin{eqnarray}
S_{\rm Simon}(u_{\rm(car)})&=&S(u_{\rm(car)})+S_a(u_{\rm(car)})+S_{\rm (lie)\theta}(u_{\rm(car)})\,,\nonumber\\
S_{\rm Simon-Mars}(u_{\rm(car)})&=&-S_a(u_{\rm(car)})\,,
\end{eqnarray}
which vanish identically for $\theta=\pi/2$.

\subsection{PG observers}

The interest for this observer family is mainly due to the fact that they can be associated with a system of horizon-penetrating coordinates. 
We have $a(u_{\rm(PG)})=0=\omega(u_{\rm(PG)})$, so that $S_a(u_{\rm(PG)})=0=S_\omega(u_{\rm(PG)})$ (and $S_z(u_{\rm(PG)})=0$ too) and 
\beq
S(u_{\rm(PG)})+S_{\rm (lie)\theta}(u_{\rm(PG)})=0\,,
\eeq
so that $S_{\rm(lie)\theta}(u_{\rm(PG)})=-S(u_{\rm(PG)})$,  with
\begin{eqnarray}\fl
S(u_{\rm(PG)})_{11}&=&-\frac{3aM\sqrt{2Mr(r^2+a^2)}}{A\Sigma^{3/2}(ir+a\cos\theta)^2}\{
\cos\theta[a^4(M-r)\sin^4\theta\nonumber\\
\fl
&&
+ra^2\sin^2\theta(-7(r^2+a^2)+2Mr)-2r(r^2+a^2)^2]\nonumber\\
\fl
&&
+ia[a^2\sin^4\theta(5r^2+4a^2+Mr)+\sin^2\theta(2a^2Mr\nonumber\\
\fl
&&
+(r^2+a^2)(5r^2-2a^2))-2(r^2+a^2)^2]
\}
\,,\nonumber\\
\fl
S(u_{\rm(PG)})_{12}&=&\frac{3aM\sqrt{2Mr(r^2+a^2)}\sin\theta}{4\sqrt{Ar}\Sigma^{3/2}(ir+a\cos\theta)^4}\{
6ra^2\sin^2\theta+r(7r^2+13a^2)\nonumber\\
\fl\qquad
&&
+ia\cos\theta(7r^2+a^2)
\}
\,,\nonumber\\
\fl
S(u_{\rm(PG)})_{13}&=&-\frac{3a^2M^2\sin\theta}{A\Sigma^{5/2}(ir+a\cos\theta)^3}\{
a\sin\theta[(r^2-a^2)a^2\sin^2\theta+(r^2+a^2)(3r^2+a^2)]\nonumber\\
\fl
&&
+2i(r^2+a^2)r\cos\theta[5a^2\sin^2\theta+3(r^2+a^2)]
\}
\,,\nonumber\\
\fl
S(u_{\rm(PG)})_{22}&=&-\frac{3aM\sqrt{2M}}{2\sqrt{r(r^2+a^2)}\Sigma^{3}(ir+a\cos\theta)^3}\{
a\sin^2\theta[(3r^2+a^2)a^2\sin^2\theta\nonumber\\
\fl\qquad
&&
+(9r^2-a^2)(r^2+a^2)]
+2ir(r^2+a^2)\cos\theta(3a^2\sin^2\theta+r^2+a^2)
\}
\,,\nonumber\\
\fl
S(u_{\rm(PG)})_{23}&=&-\frac{\sqrt{2Mr}a\sin\theta}{\sqrt{\Sigma(r^2+a^2)}}S(u_{\rm(PG)})_{12}
\,,\nonumber\\
\fl
S(u_{\rm(PG)})_{33}&=&-\frac{3aM\sqrt{2M}}{A\sqrt{r(r^2+a^2)}\Sigma^{3}(ir+a\cos\theta)^3}\{
a\sin^2\theta[-(4Mr^3-(r^2+a^2)^2)a^4\sin^4\theta\nonumber\\
\fl
&&
-2a^2(r^2+a^2)(a^4-r^4+6Mr^3)\sin^2\theta+(a^2-3r^2)(r^2+a^2)^3]\nonumber\\
\fl
&&
+2ir(r^2+a^2)\cos\theta[-a^4\sin^4\theta(r^2+a^2+8Mr)-4Mra^2(r^2+a^2)\sin^2\theta\nonumber\\
\fl
&&
+(r^2+a^2)^3]
\}
\,.
\end{eqnarray}
Note that in this case 
\beq
S_{\rm Simon}(u_{\rm(PG)})=0=S_{\rm Simon-Mars}(u_{\rm(PG)})\,,
\eeq
identically. 
Furthermore, the spatial field $z(u_{\rm(PG)})\equiv0$, being the PG congruence both geodesic and irrotational.

\section{Concluding remarks}

The Simon and Simon-Mars tensors have been introduced to characterize the Kerr spacetime within certain classes of stationary asymptotic flat solutions of the vacuum Einstein's field equations admitting a timelike Killing vector field \cite{Simon,krisch88,Mars99}. Their vanishing with respect to this preferred observer family (the static observers) has been explained in terms of the alignment of the principal null directions of the Weyl tensor of the spacetime with those of the Papapetrou field associated with the Killing congruence \cite{Fasop99,Mars00,Fasop00}.
In recent years there have been a renewed interest in the Simon-Mars tensor as a mathematical tool to investigate the geometrical properties of more general stationary spacetimes and to classify them \cite{Mars:2016vtr,Mars:2016ahh,Mars:2016ynw,Paetz:2017lkn}. Furthermore, estimating the nonvanishing of the the Simon-Mars components provides and indication of the deviation of a given stationary spacetime (e.g., generated by a source with multipolar structure) from a Kerr-like behavior, at least locally \cite{Some:2014kfa}.  

We have generalized the definition of Simon tensor, originally given only in the Kerr spacetime and only associated with the static (Killing) family of observers, to any spacetime and to any observer.
We have accomplished this generalization by a standard \lq\lq 3+1" splitting of the divergence equation of the Weyl tensor (equivalent to the Bianchi identities).
We have shown that the newly defined relative-observer Simon tensor can be used to write a  \lq\lq balance equation" among various fields, associated with the kinematical properties of the observer congruence and representing (pieces of) the spacetime curvature in general.

The usefulness of this relative-observer definition of the Simon tensor is at least twofold. It is shown to naturally arise by splitting the Bianchi identities of the first kind for any spacetime and any observer family, in  the sense that it comes automatically when exploring the $3+1$ content of the Bianchi identities.
It contains relative-observer curvature information encoded in three spatial fields associated with the kinematical properties of the splitting congruence. 
A change of observer thus results in transferring curvature information from one field to the other.
Explicit examples considered for the case of a Kerr black hole spacetime illuminate once more the relative-observer physics.
We have examined the well known families of ZAMOs and Carter observers as well as the less familiar Painlev\'e-Gullstrand family of observers. 
The latter plays an important role in this context. In fact, being geodesic and irrotational, it has associated an identically vanishing Simon-Mars tensor without any alignment property of principal null directions, differently from the static observer case. 
Carter observers are once more special, since the various fields decomposing the Simon and Simon-Mars tensors are all proportional, just as the electric and magnetic parts of the Weyl tensor, and vanish identically on the equatorial plane.
Finally, ZAMOs never measure a zero Simon-Mars tensor. 

The present analysis can be applied to non-vacuum spacetimes as well. We have considered a simple generalization of the Kerr solution by the inclusion of a cosmological constant term, i.e., the Kerr-de Sitter spacetime, belonging to a family of solutions whose characterization in terms of the Simon-Mars tensor has been recently investigated in Refs. \cite{Mars:2016vtr,Mars:2016ahh,Mars:2016ynw,Paetz:2017lkn}. We have found that the associated Cotton gravitational current vanishes, so that the Simon and Simon-Mars tensors admit the same decomposition as in the Kerr case. Furthermore, the presence of a cosmological constant allows the Simon-Mars tensor to vanish as measured by ZAMOs.

\appendix

\section{Kerr spacetime: properties of the static, ZAMO, Carter, PG observers}

We list below the relevant kinematical quantities as well as the electric and magnetic part of the Weyl tensor for the various families of observers.

\subsection{Static observers}

The static observers are accelerated, with 4-acceleration
\begin{eqnarray}
a(m)&=&\frac{M\sqrt{\Delta}(r^2-a^2\cos^2\theta)}{\Sigma^{3/2}(\Sigma-2Mr)}e_{\hat r}
-\frac{2Mra^2 \sin \theta \cos \theta}{\Sigma^{3/2}(\Sigma-2Mr)}e_{\hat \theta}\,,
\end{eqnarray}
and are locally rotating, with vorticity vector
\begin{eqnarray}
\omega(m)&=&-\frac{2aMr \sqrt{\Delta}\cos\theta}{\Sigma^{3/2}(\Sigma-2Mr)} e_{\hat r}
 -\frac{Ma(r^2-a^2\cos^2\theta)\sin \theta}{\Sigma^{3/2}(\Sigma-2Mr)}e_{\hat \theta}\,,
\end{eqnarray}
but the  congruence of their world lines is not expanding, i.e., has vanishing expansion $\theta(m)=0$. 
The components of the electric part of the Weyl tensor are given by
\begin{eqnarray}
\label{variousqua_thd}
E(m)_{11}&=&-\frac{Mr}{\Sigma^3}(4r^2-3\Sigma)  \frac{2\Delta+a^2\sin^2\theta}{\Sigma-2Mr}\ , \nonumber\\
E(m)_{12}&=&\frac{3Ma^2\sqrt{\Delta}}{\Sigma^3}(4r^2-\Sigma)\frac{\cos\theta\sin\theta}{\Sigma-2Mr}\ , \nonumber\\
E(m)_{22}&=&\frac{Mr}{\Sigma^3}(4r^2-3\Sigma)\frac{\Delta+2a^2\sin^2\theta}{\Sigma-2Mr}\ , \nonumber\\
E(m)_{33}&=&\frac{Mr}{\Sigma^3}(4r^2-3\Sigma)\ , 
\end{eqnarray}
and 
\begin{eqnarray}
&&H(m)_{11}=kE(m)_{11}\ , \quad 
H(m)_{12}=-k^{-1}E(m)_{12}\ , \nonumber\\ 
&&
H(m)_{22}=kE(m)_{22}\ , \quad
H(m)_{33}=kE(m)_{33}\ , 
\end{eqnarray}
with
\beq
\label{kdef}
k=\frac{a}{r}\frac{4r^2-\Sigma}{4r^2-3\Sigma}\cos\theta\,.
\eeq
The quantity $k$ entering this relation (and the analogous below) is simply related to the expression of the curvature 2-forms in the Carter (principal) frame of the spacetime. Indeed, it is given by $k=J/I$, where
\beq
J=\frac{Ma\cos\theta}{\Sigma^3}(3r^2-a^2\cos^2\theta)\,,\quad
I=\frac{Mr}{\Sigma^3}(r^2-3a^2\cos^2\theta)\,,
\eeq
see, e.g., Ref. \cite{Oneill}, Sec. 2.7, p. 96.

Finally, let us note the important property 
\beq
[\nabla(m)z(m)]^{\rm STF}=\kappa [z(m)\otimes z(m)]^{\rm STF}\,,
\eeq
with 
\beq
\kappa=1-\frac{3i(a\cos\theta+ir)}{M}+\frac{6ir}{a\cos\theta-ir}\,.
\eeq

\subsection{ZAMOs}

The accelerated ZAMOs are locally nonrotating in the sense that their vorticity vector $\omega(n)$ vanishes, but they have a nonzero expansion tensor $\theta(n)$ with vanishing expansion scalar. 

We list below the nonvanishing components of the kinematical fields:
acceleration
\begin{eqnarray}\fl\quad
a(n)_{1}&=&-\frac{M}{\sqrt{\Delta}\Sigma^{3/2}A}
\left\{a^2\cos^2\theta[(r^2+a^2)^2-4Mr^3]-r^2[(r^2+a^2)^2-4a^2Mr]\right\}\,,
\nonumber \\
\fl\quad
a(n)_{2}&=&-\frac{ 2\sin\theta\cos\theta Mr a^2 (r^2+a^2)}{\Sigma^{3/2}A} 
\,,
\end{eqnarray}
and expansion tensor
\begin{eqnarray}
\theta(n)_{13}&=& -\frac{aM\sin\theta}{\Sigma^{3/2}A}[r^2(3r^2+a^2)
+a^2(r^2-a^2)\cos^2\theta)]
\,,\nonumber \\
\theta(n)_{23}&=&\frac{2ra^3M\sin^2\theta\cos\theta\sqrt{\Delta}}{\Sigma^{3/2}A}\,.
\end{eqnarray}

The components of the electric part of the Weyl tensor are given by
\begin{eqnarray}
\label{variousqua_zamo}
E(n)_{11}&=&-\frac{Mr}{A\Sigma^3}(4r^2-3\Sigma)[2(r^2+a^2)^2+a^2\Delta\sin^2\theta]\ , \nonumber\\
E(n)_{12}&=&\frac{3a^2M\sqrt{\Delta}}{A\Sigma^3}(4r^2-\Sigma)(r^2+a^2)\cos\theta\sin\theta\ , \nonumber\\
E(n)_{22}&=&\frac{Mr}{A\Sigma^3}(4r^2-3\Sigma)[(r^2+a^2)^2+2a^2\Delta\sin^2\theta]\ , \nonumber\\
E(n)_{33}&=&\frac{Mr}{\Sigma^3}(4r^2-3\Sigma)\ , 
\end{eqnarray}
and 
\begin{eqnarray}
&&H(n)_{11}=kE(n)_{11}\ , \quad 
H(n)_{12}=-k^{-1}E(n)_{12}\ , \nonumber\\ 
&&
H(n)_{22}=kE(n)_{22}\ , \quad
H(n)_{33}=kE(n)_{33}\ , 
\end{eqnarray}
with $k$ given in Eq. (\ref{kdef}).

\subsection{Carter observers}

The Carter observers are accelerated, with 4-acceleration
\begin{eqnarray}\fl\quad
a(u_{\rm (car)})&=&\frac{M(r^2-a^2\cos^2\theta)+ra^2\sin^2\theta}{\sqrt{\Delta}\Sigma^{3/2}}e(u_{\rm (car)})_1 
-\frac{a^2 \sin \theta \cos \theta}{\Sigma^{3/2}}e(u_{\rm (car)})_2\,,
\end{eqnarray}
and are locally rotating, with vorticity vector
\begin{eqnarray}
\omega(u_{\rm (car)})&=&\frac{a\sqrt{\Delta}\cos\theta}{\Sigma^{3/2}} e(u_{\rm (car)})_1\,,
\end{eqnarray}
and expanding, with the following only nonvanishing frame component of the expansion tensor
\beq
\theta(u_{\rm (car)})_{13}=-\frac{ar \sin \theta}{\Sigma^{3/2}}\,.
\eeq
The components of the electric part of the Weyl tensor are given by
\begin{eqnarray}
\label{variousqua_car}
\fl\qquad
E(u_{\rm (car)})&=&\frac{Mr}{\Sigma^3}(4r^2-3\Sigma)\,
{\rm diag}[-2,1,1]=k^{-1}\,H(u_{\rm (car)})_{ab}\ , 
\end{eqnarray}
with $k$ given by Eq. (\ref{kdef}).

\subsection{Painlev\'e-Gullstrand observers}
\label{appKPG}

The kinematical properties of the $u_{\rm(PG)}$ observers are summarized by the expansion
$\theta(u_{\rm(PG)})$. In fact the exterior derivative of (\ref{PG4vel}) is zero: $du_{\rm(PG)}=0$, implying that $a(u_{\rm(PG)})=0$ and $\omega(u_{\rm(PG)})=0$.
The nonvanishing frame  components of the expansion tensor are given by
\begin{eqnarray}\fl
\label{thetaPG}
\theta(u_{\rm(PG)})_{11}&=&\frac{\sqrt{2Mr}}{2rA\Sigma(r^2+a^2)^{1/2}}[\Sigma(r^4-a^4)+2r^2a^2\Delta\sin^2\theta]\ , \nonumber\\
\fl
\theta(u_{\rm(PG)})_{12}&=&-\frac{a^2\sqrt{2Mr(r^2+a^2)}}{A\Sigma^{3/2}}\sin\theta\cos\theta\ , \nonumber\\
\fl
\theta(u_{\rm(PG)})_{13}&=&-\frac{aM}{A\Sigma^{3/2}}[(r^2-a^2)\Sigma+2r^2(r^2+a^2)]\sin\theta\ \ , \nonumber\\
\fl
\theta(u_{\rm(PG)})_{22}&=&-\frac{r\sqrt{2Mr(r^2+a^2)}}{\Sigma^2}\ , \nonumber\\
\fl
\theta(u_{\rm(PG)})_{23}&=&\frac{2a^3Mr}{\sqrt{A}\Sigma^2}\cos\theta\sin^2\theta\ , \nonumber\\
\fl
\theta(u_{\rm(PG)})_{33}&=&-\frac{\sqrt{2Mr(r^2+a^2)}}{A\Sigma^2}[(r-M)\Sigma^2+M(3r^2+a^2)\Sigma-2Mr^2(r^2+a^2)]\ .
\end{eqnarray}

The components of the electric part of the Weyl tensor are listed below:
\begin{eqnarray}\fl\quad
E(u_{\rm(PG)})_{11}&=&-\frac{Mr}{A\Sigma^3}(4r^2-3\Sigma)[2(r^2+a^2)^2+a^2\Delta\sin^2\theta]\ , \nonumber\\
\fl\quad
E(u_{\rm(PG)})_{12}&=&\frac{3Ma^2}{\sqrt{A}\Sigma^{7/2}}(4r^2-\Sigma)(r^2+a^2)\cos\theta\sin\theta\ , \nonumber\\
\fl\quad
E(u_{\rm(PG)})_{13}&=&\frac{3aMr\sqrt{2Mr}(r^2+a^2)^{3/2}}{A\Sigma^{7/2}}(4r^2-3\Sigma)\sin\theta\ , \nonumber\\
\fl\quad
E(u_{\rm(PG)})_{22}&=&\frac{Mr}{\Sigma^4}(4r^2-3\Sigma)[3(r^2+a^2)-2\Sigma]\ , \nonumber\\
\fl\quad
E(u_{\rm(PG)})_{23}&=&-\frac{3a^3M\sqrt{2Mr(r^2+a^2)}}{\sqrt{A}\Sigma^4}(4r^2-\Sigma)\cos\theta\sin^2\theta\ , \nonumber\\
\fl\quad
E(u_{\rm(PG)})_{33}&=&-\frac{Mr}{A\Sigma^4}(4r^2-3\Sigma)\{-\Sigma^2\Delta+2Mr[3(r^2+a^2)-4\Sigma]\}\ ,
\end{eqnarray}
and 
\begin{eqnarray}
\fl\quad
&&H(u_{\rm(PG)})_{11}=kE(u_{\rm(PG)})_{11}\ , \quad 
H(u_{\rm(PG)})_{12}=-k^{-1}E(u_{\rm(PG)})_{12}\ , \nonumber\\ 
\fl\quad
&&H(u_{\rm(PG)})_{13}=kE(u_{\rm(PG)})_{13}\ , \quad 
H(u_{\rm(PG)})_{22}=kE(u_{\rm(PG)})_{22}\ , \nonumber\\ 
\fl\quad
&&
H(u_{\rm(PG)})_{23}=-k^{-1}E(u_{\rm(PG)})_{23}\ , \quad 
H(u_{\rm(PG)})_{33}=kE(u_{\rm(PG)})_{33}\ , 
\end{eqnarray}
with $k$ given by Eq. (\ref{kdef}).

\section{Non-vacuum solutions: the case of a Kerr-de Sitter spacetime}

Examples of non-vacuum spacetimes have been considered, e.g., in Refs. \cite{Kramer85,Mars:2001ge,Some:2014kfa,Bini:2004qf,Bini:2004sg}, including electrovacuum solutions and perfect fluid solutions, for static observers only.
We will discuss below the case of a Kerr-de Sitter spacetime, which is a generalization of the Kerr solution to account for the presence of a cosmological constant term into the Einstein's equations, with associated energy momentum tensor $T_{\mu\nu}=\Lambda g_{\mu\nu}$ (see, e.g., Ref. \cite{Akcay:2010vt} and references therein).
Its line element written in Boyer-Lindquist-like coordinates $(t,r,\theta,\phi)$ is given by
\begin{eqnarray}\fl\qquad
ds^2 &=& -\frac{\Delta_r}{\Sigma k^2}\left[1-\frac{\Delta_\theta}{\Delta_r}a^2\sin^2\theta\right]dt^2 
-\frac{2a}{\Sigma k^2}[(r^2+a^2)\Delta_\theta-\Delta_r]\sin^2\theta dtd\phi\nonumber\\
\fl\qquad
&&
+\frac{\Sigma}{\Delta_r}dr^2
+\frac{\Sigma}{\Delta_\theta} d\theta^2+\frac{\mathcal A}{\Sigma k^2}\sin^2 \theta d\phi^2\,,
\end{eqnarray}
with 
\begin{eqnarray}
\Delta_r&=&(r^2+a^2)\left(1-\frac{\Lambda}{3}r^2\right)-2Mr
=\Delta-\frac{\Lambda}{3}r^2(r^2+a^2)\,,\nonumber\\
\Delta_\theta&=&1+\frac{\Lambda}{3}a^2\cos^2\theta\,,\nonumber\\
{\mathcal A}&=&(r^2+a^2)^2\Delta_\theta-\Delta_r a^2\sin^2\theta
=A+\frac{\Lambda}{3}a^2(r^2+a^2)\Sigma\,,
\end{eqnarray}
and $k=1+\Lambda a^2/3$.
The limiting case of a Kerr black hole ($\Lambda=0$) is recovered by replacing $\Delta_r\to\Delta$, $\Delta_\theta\to1$, ${\mathcal A}\to A$ and $k\to1$.

The associated Cotton gravitational current $J^{\rm(G)}=0$, so that the decomposition of the Simon tensor is still given by Eq. (\ref{simontens}) as in the Kerr case.
We will consider below the families of static observers and ZAMOs.

\subsection{Static observers}

The static observers have 4-velocity
\beq
m=\frac1{\sqrt{-g_{tt}}}\,\partial_t\,.
\eeq
An orthonormal frame adapted to $m$ is
\begin{eqnarray}\fl\qquad
e(m)_1 &=&
\frac1{\sqrt{g_{rr}}}\,\partial_r
\,,\qquad
e(m)_2 = 
\frac1{\sqrt{g_{\theta\theta}}}\,\partial_\theta
\,,\nonumber\\
\fl\qquad
e(m)_3 &=&
\left[\frac{k^2(\Delta_r-\Delta_\theta a^2\sin^2\theta)}{\sin^2 \theta \Delta_r\Delta_\theta \Sigma}\right]^{1/2} \left(\partial_\phi-a\sin^2\theta\frac{(r^2+a^2)\Delta_\theta-\Delta_r}{\Delta_r-\Delta_\theta a^2\sin^2\theta} \partial_t  \right)
\,. \nonumber\\
\end{eqnarray}
The Simon and Simon-Mars tensors both vanish in this case, since $S_a(m)=-S(m)$, $S_\omega(m)=iS(m)$ and $S_z(m)=0$.
The nonvanishing frame components of $S(m)$ are given by
\begin{eqnarray}\fl
S(m)_{13}&=&-\frac32aM^2\sin\theta\Delta_r\sqrt{\Delta_\theta}\frac{ir-a\cos\theta}{[\Delta_r-\Delta_\theta a^2\sin^2\theta]^2\Sigma^{5/2}}
\left[1-i\frac{\Lambda}{3M}(ir-a\cos\theta)^3\right]\,,\nonumber\\
\fl
S(m)_{23}&=&ia\sin\theta\left(\frac{\Delta_\theta}{\Delta_r}\right)^{1/2}S(m)_{13}\,.
\end{eqnarray}

\subsection{ZAMOs}

ZAMOs have 4-velocity 
\beq
n=N^{-1}(\partial_t-N^{\phi}\partial_\phi)\,,
\eeq
where 
\beq\fl\quad
N=(-g^{tt})^{-1/2}=\left[\frac{\Delta_r\Delta_\theta\Sigma}{k^2\mathcal A}\right]^{1/2}\,,\qquad
N^{\phi}=\frac{g_{t\phi}}{g_{\phi\phi}}=-\frac{a}{\mathcal A}[(r^2+a^2)\Delta_\theta-\Delta_r]
\eeq
are the lapse and shift functions, respectively.
An orthonormal frame adapted to $n$ is
\begin{eqnarray}
e(n)_1&=&\frac1{\sqrt{g_{rr}}}\,\partial_r\,, \quad
e(n)_2=\frac1{\sqrt{g_{\theta \theta }}}\,\partial_\theta\,, \quad
e(n)_3=\frac1{\sqrt{g_{\phi \phi }}}\,\partial_\phi\,.
\end{eqnarray}
In this case 
\beq\fl\qquad
S_{\rm Simon}(n)=S(n)+S_a(n)+S_{\rm (lie)\theta}(n)\,,\qquad
S_{\rm Simon-Mars}(n)=-S_a(n)\,.
\eeq
The nonvanishing components of $S(n)$, $S_a(n)$ and $S_{\rm (lie)\theta}(n)$ are listed below
\begin{eqnarray}\fl\qquad
[S_a(n)]_{13}&=&\sqrt{\Delta_\theta}\frac{A^2}{{\mathcal A}^2}\left[
[S_a(n)]_{13}^{\rm Kerr}
+\Lambda(B_1+B_2\Lambda)\frac{(r^2+a^2)^2aM\sin\theta}{2A^2\Sigma^{3/2}(ir+a\cos\theta)^3}
\right]\,,\nonumber\\
\fl\qquad
{}[S_a(n)]_{23}&=&\frac{ia\sin\theta}{r^2+a^2}\left(\frac{\Delta_r}{\Delta_\theta}\right)^{1/2} [S_a(n)]_{13}\,,
\end{eqnarray}
with
\begin{eqnarray}\fl\qquad
B_1&=&r^4[r(r^2+a^2)+4Ma^2]+r^2[(3a^2-5r^2)M+2r(r^2+a^2)]a^2\cos^2\theta\nonumber\\
\fl\qquad
&&
+[M(a^2-3r^2)+r(r^2+a^2)]a^4\cos^4\theta\nonumber\\
\fl\qquad
&&
+ia\cos\theta[r(r^2+2a^2\cos^2\theta)(r^3+ra^2+2Ma^2)+\Delta a^4\cos^4\theta]
\,,\nonumber\\
\fl\qquad
B_2&=& \frac13a^2(r^2+a^2)\Sigma^2(r+ia\cos\theta)\,,
\end{eqnarray}
and
\begin{eqnarray}\fl\qquad
[S_{\rm(lie)\theta}(n)]_{13}&=&\sqrt{\Delta_\theta}\frac{A^2}{{\mathcal A}^2}\left[
[S_{\rm(lie)\theta}(n)]_{13}^{\rm Kerr}
-\Lambda\frac{C_1(r^2+a^2)a^3M^2\sin\theta}{2A^2\Sigma^{3/2}(ir+a\cos\theta)^3}
\right]\,,\nonumber\\
\fl\qquad
{}[S_{\rm(lie)\theta}(n)]_{23}&=&\left(\frac{\Delta_r}{\Delta}\right)^{1/2}\frac{A^2}{{\mathcal A}^2}\left[
[S_{\rm(lie)\theta}(n)]_{23}^{\rm Kerr}\right.\nonumber\\
\fl\qquad
&&\left.
-\Lambda\frac{C_2\sqrt{\Delta}(r^2+a^2)a^4M^2\cos\theta\sin^2\theta}{2A^2\Sigma^{3/2}(ir+a\cos\theta)^3}
\right]\,,
\end{eqnarray}
with
\begin{eqnarray}\fl\qquad
C_1&=&[2r^2-(5r^2+3a^2)\cos^2\theta][(r^2-a^2)\Sigma+2r^2(r^2+a^2)]\nonumber\\
\fl\qquad
&&
-10iar^3(r^2+a^2)\cos\theta\sin^2\theta
\,,\nonumber\\
\fl\qquad
C_2&=&2ar[3r^2-(5r^2+2a^2)\cos^2\theta]-5i\cos\theta[(r^2-a^2)\Sigma+2r^2(r^2+a^2)]
\,,
\end{eqnarray}
and
\begin{eqnarray}\fl\qquad
S(n)_{13}&=&\sqrt{\Delta_\theta}\frac{A^2}{{\mathcal A}^2}\left[
S(n)_{13}^{\rm Kerr}
-\Lambda(D_1+D_2\Lambda)\frac{(r^2+a^2)aM\sin\theta}{2A^2\Sigma^{3/2}(ir+a\cos\theta)^3}
\right]\,,\nonumber\\
\fl\qquad
S(n)_{23}&=&\left(\frac{\Delta_r}{\Delta}\right)^{1/2}\frac{A^2}{{\mathcal A}^2}\left[
S(n)_{23}^{\rm Kerr}
+\Lambda(D_3+D_4\Lambda)\frac{\sqrt{\Delta}(r^2+a^2)a^2M\sin^2\theta}{2A^2\Sigma^{3/2}(ir+a\cos\theta)^3}
\right]\,,\nonumber\\
\fl\qquad
\end{eqnarray}
with
\begin{eqnarray}\fl\qquad
D_1&=&[(2r-M)(r^2+a^2)^2-4a^2Mr^2]\Sigma^2+Mr^2(7r^2+13a^2)(r^2+a^2)\Sigma\nonumber\\
\fl\qquad
&&
-6Mr^4(r^2+a^2)^2\nonumber\\
\fl\qquad
&&
+2ia(r^2+a^2)\cos\theta[\Delta\Sigma^2-Mr(r^2-4a^2)\Sigma+3Mr^3(r^2+a^2)]
\,,\nonumber\\
\fl\qquad
D_2&=&-2iB_2\,,\nonumber\\
\fl\qquad
D_3&=&2a\cos\theta[\Delta\Sigma^2-Mr(r^2-2a^2)\Sigma+3Mr^3(r^2+a^2)]\nonumber\\
\fl\qquad
&&
-i\{[2r(r^2+a^2)-M(r^2+3a^2)]\Sigma^2+Mr^2(7r^2+17a^2)\Sigma\nonumber\\
\fl\qquad
&&
-6Mr^4(r^2+a^2)\}\,,\nonumber\\
\fl\qquad
D_4&=&i(r^2+a^2)D_2\,.
\end{eqnarray}

\begin{figure}
\centering
\includegraphics[scale=0.35]{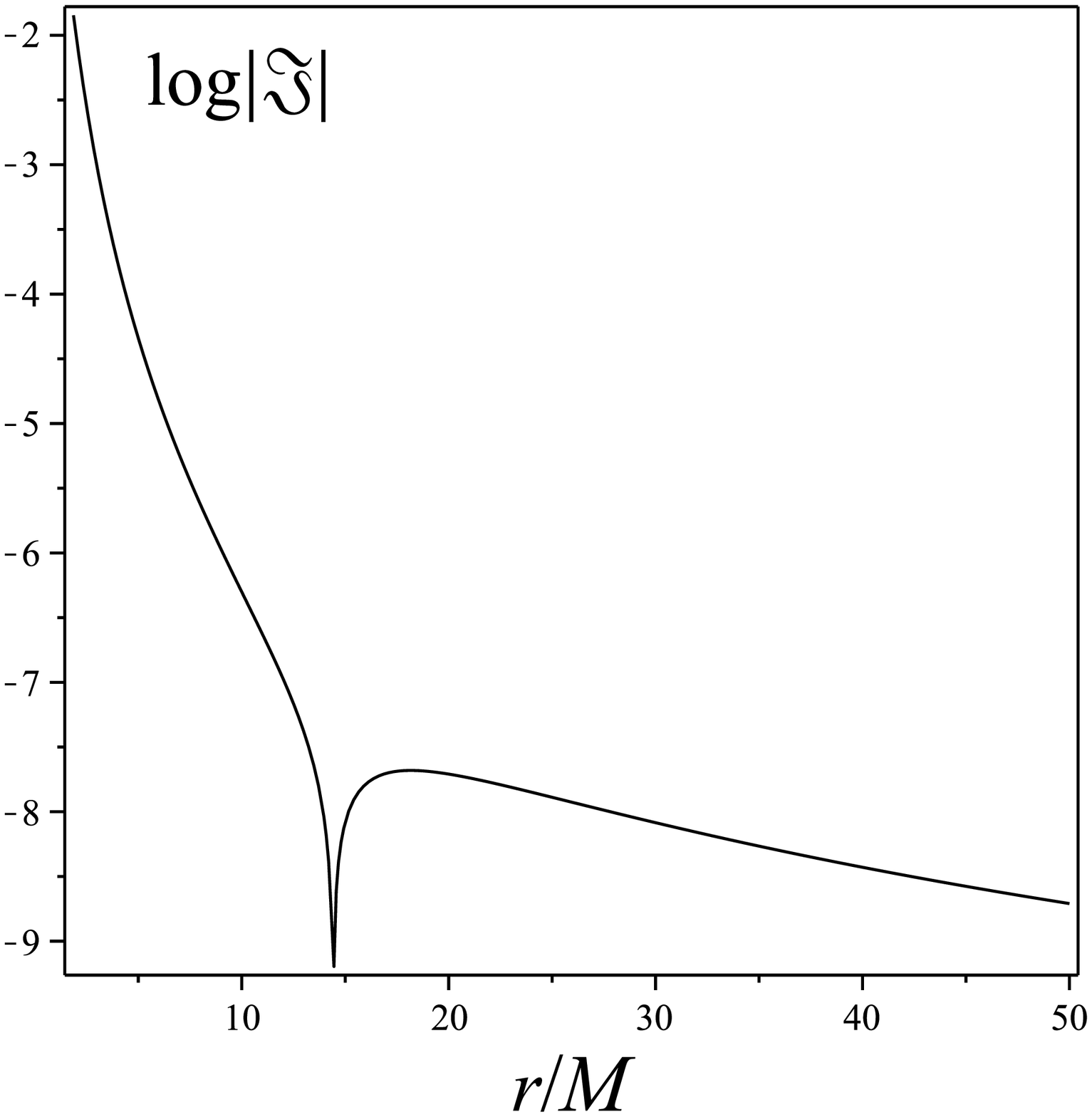}
\caption{\label{fig:2} 
The behavior of the (log10 of) the absolute value of the imaginary part $\mathcal{J}$ of the frame component $[S_a(n)]_{13}$ is shown as a function of the radial coordinate for $\theta=\pi/2$, $a/M=0.5$ and $M^2\Lambda=10^{-3}$.
It vanishes at $r/M\approx14.42$, implying that ZAMOs measure zero Simon-Mars tensor there. 
}
\end{figure}

In contrast with the Kerr case, ZAMOs can measure a vanishing Simon-Mars tensor, since the only independent frame component $[S_a(n)]_{13}=0$ for $\theta=\pi/2$ and a value of the radial coordinate given by the condition ${\rm Im}([S_a(n)]_{13})=0$ (see Fig. \ref{fig:2}).


\section*{Acknowledgements}

D.B. thanks ICRANet and the Italian Istituto Nazionale di Fisica Nucleare (INFN) for partial support.

\end{document}